\documentclass{amsart}
\usepackage{amssymb,amsthm,euscript,amsmath,graphicx}
\usepackage[english]{babel}
\usepackage{t1enc}

\newcommand{\be}{\begin{equation}}
\newcommand{\ee}{\end{equation}}
\newcommand{\ben}{\begin{equation*}}
\newcommand{\een}{\end{equation*}}
\newcommand{\bg}{\begin{gather}}
\newcommand{\eg}{\end{gather}}
\newcommand{\bea}{\begin{eqnarray}}
\newcommand{\eea}{\end{eqnarray}}
\newcommand{\eean}{\end{eqnarray*}}
\newcommand{\bean}{\begin{eqnarray*}}
\newcommand\re[1]{(\ref{#1})}

\newcommand{\bsq}{\hbox{\vrule width 3pt height 3pt depth 0pt}}

\begin{document}

\title{On the continuity and Lesche stability of Tsallis and Rényi entropies and q-expectation values}
\author{Matolcsi$^1$ T. and  V\'an$^{23}$ P.}
\address{$^1$Department of Applied Analysis and Computational Mathematics\\ 
        E\"otv\"os Lor\'and University, Budapest, Hungary\\
$^2$Department of Theoretical Physics\\
        KFKI, Research Institute of Particle and Nuclear Physics, Budapest,
        Hungary\\
$^3$Department of Energy Engineering\\
        Budapest University of Technology and Economics, Hungary}

\email{vpet@rmki.kfki.hu}

\date{\today}
\begin{abstract}

It is shown that the Rényi and Tsallis entropies and the q-expectation values,
are continuous and stable if $q>1$  and are not continuous and instable for
uniform finite distributions if
$q<1$. 
\end{abstract}
\maketitle


\section{Introduction}

\textit{Experimental robustness} is a natural criteria of physical quantities
requiring that  \ 

\begin{quote} A physically meaningful function of a probability
distribution should not change drastically if the underlying distribution
function is slightly changed. 
\end{quote} 

Lesche in 1982 has given a mathematical formulation of the above requirement,
called stability, and proved that the entropy of Rényi is not stable  \cite{Les82a}. 
Based on Lesche's reasoning later on Abe has shown that the 
Tsallis entropy is stable \cite{Abe02a}. Lesche stability
became a  criteria in distinguishing and favoring one of the many different
entropies in non-extensive thermostatistics  \cite{AbeAta04a,Abe03a,TsaBri03a,KanSca04a,KanAta05a,Bec09a,Kan09a}
and the proofs of
Lesche and  Abe become one of the arguments in favoring Tsallis
entropy to Rényi. Lesche stability as a proper concept of experimental robustness
was questioned and attacked by several authors
\cite{JizAri04a1,Bas05a,Yam04a}. They have collected physical arguments claiming
that Lesche stability do not express properly the physical content of experimental
robustness. Lesche and Abe rejected
these arguments \cite{Les04a,Abe05a}.
Recently Abe recognized that the central quantities of non-extensive statistical
mechanics, the q-averages \cite{AbeBag05a}, are Lesche instable \cite{Abe08a}. This important observation
somehow invalidates the whole mathematical framework of non-extensive thermostatistics,
therefore several authors argued again that Lesche stability is
a too strict concept for physical applications and suggested different modifications \cite{HanAta09a,LutAta09a,Abe09a}.

The concept of experimental robustness is a lousy continuity requirement
and enables several mathematical formulations. Considering this resemblance
to continuity the instability of the Rényi entropy \(S_R\) \re{SR} and the 
stability of Tsallis entropy \(S_T\) \re{ST} is somehow paradoxical, because 
the Tsallis entropy \(S_T=(1-e^{(1-q)S_R})/(q-1)\) (where  \(0<q\neq1)\) is a 
smooth function of the Rényi entropy. 

In the following we investigate some mathematical concepts releted to
the Rényi and Tsallis entropies and q-expectation values. We introduce a 
local form of Lesche stability, that, according to our opinion, expresses 
best the physical content of experimental robustness. 

\section{Continuity of functions of probability distributions}

The simplest formulation of experimental robustness is continuity. 
Recall the following notions. 

The set of infinite discrete probability distributions is 
$$
D := \left\{ p\in l^1 |\  \|p\|_1 = 1,\ p_i\geq 0,\ i\in \mathbb{N}\right\}
\subset l^1.
$$

Here the $l^1$ norm is used as the natural concept of distance \cite{AbeAta07a}.

Let $X$ be a normed space with norm $\| \ \|$. 

\textit{Definition 1:} A function   $f:D\rightarrow X$ is \textit{continuous at $p$} if 
$$
(\forall \epsilon>0)(\exists \delta>0)(\forall r)(\|r-p\|_1 <\delta \Rightarrow\|f(r)-f(p)\|
< \epsilon).  
$$

$f$ is \emph{continuous} if it is continuous at every $p\in D$.   

Note that if there  is a positive number $c_p$ so that $\|f(r)-f(p)\| < c_p\|r-p\|_1$
then $f$ is continuous at $p$. 

\textit{Definition 2:} A function   $f:D\rightarrow X$ is \textit{uniformly 
continuous} if 
$$
(\forall \epsilon>0)(\exists \delta>0)(\forall r,p)(\|r-p\|_1 <\delta \Rightarrow\|f(r)-f(p)\|
< \epsilon).  
$$

Note that if there  is a positive number $c$ so that $\|f(r)-f(p)\| < c\|r-p\|_1$
then $f$ is uniformly continuous. 

Continuity is a \emph{local} property while uniform continuity is a \emph{global}
property.

Observe that the negation of continuity reads as follows:
$$
(\exists p)(\exists \epsilon>0)(\forall \delta>0)(\exists r, \|r - p\|_1 <\delta )(\|f(r)-f(p)\|
\geq\ \epsilon)  
$$
and the negation of unformly continuity:
$$
(\exists \epsilon>0)(\forall \delta>0)(\exists r,p, \|r - p\|_1 <\delta )
(\|f(r)-f(p)\|\geq\ \epsilon).
$$

\subsection{1<q}\label{stab1}

The Banach space of real sequences for which of the corresponding
series is convergent at the power $q$, is denoted
by $l^q$, and the Banach space of bounded sequences is denoted  by $l^\infty$.
We know that if $k\in l^1$ and $1< q$, then $k\in l^q$ and $\|k\|_q \leq \|k\|_1$. Therefore
the q-norm, as the function $\|.\|_q:\ l^1\rightarrow\mathbb{R}, 
k\mapsto\|k\|_q$ function, is uniformly continuous.

{\bf Proposition 1} The function $D\rightarrow l^1, \ p\mapsto p^q:=
(p_i^q)_{i\in\mathbb{N}}$ is uniformly continuous. 

{\em Proof}: According to the mean value theorem of differential calculus

$$
\|r^q-p^q\|_1 = \sum_{i\in\mathbb{N}} |r^q_i-p^q_i| \leq 
 \sum_{i\in\mathbb{N}}q |r_i-p_i| = q \|r-p\|_1.
$$
\bsq

Note that $\|p^q\|_1=\|p\|_q$.

{\bf Corollary 1.1} The Rényi entropy  
\begin{equation}
S_{R}:\ D\rightarrow \mathbb{R}, \qquad p\mapsto \frac{1}{1-q} \ln \|p\|_q,
\label{SR}\end{equation}
is continuous and the Tsallis entropy
\begin{equation}
S_{T}:\ D\rightarrow \mathbb{R}, \qquad p\mapsto \frac{1-\|p\|_q}{q-1}
\label{ST}\end{equation}
if $ 1<q$ is uniformly continuous. \bsq

The \textit{expectation value} of $A=(A_i)_{i\in\mathbb{N}}\in l^\infty$,
$$
\ D\rightarrow \mathbb{R}, \qquad p\mapsto (A|p)=\sum_{i\in\mathbb{N}} A_ip_i,
$$  
is uniformly continuous.

In general, if $\Phi: D\rightarrow D$ is a given function, then the
$\Phi$-expectation value of $A$ is
$$
D\rightarrow \mathbb{R}, \qquad p\mapsto (A|\Phi(p)).
$$  

If $\Phi$ is (uniformly) continuous, then the $\Phi$-expectation value is 
(uniformly) continuous.

{\bf Corollary 1.2} The q-expectation value, where 
$\Phi(p)_i:= \frac{p_i^q}{\|p^q\|_{1}}$ 
(the quotient of continuous functions) is continuous.

\subsection{q<1 }\label{nstab1}
 
In this case the summability of $p^q$ for $p\in D$ is not automatic. Therefore
the previous functions (entropies and q averages) are interpreted on the
set:\ $$
 D_q := \{p\in D |\ \ p^q \in l^1\}.
$$

{\bf Proposition 2} The function \(D_q\rightarrow l^1, \ p\mapsto p^q\) is not continuous
at finite uniform distributions.

\textit{Proof:}  Let be $n\in \mathbb{N}$ a given number and 
$$
p:= \left(\frac{1}{n}, ..., \frac{1}{n},0,0,... \right)\in D,
$$
therefore the number of nonzero elements is $n$. In the following we will
use the notation
\begin{equation}
p=\left(\left.\frac{1}{n}\right|_{\times n}, 0\right).
\label{d1}\end{equation}
 
For all $0<\delta<1/2$ let us define   
\begin{equation}
r_\delta:=\left(\left.\frac{1-\delta}{n}\right|_{\times n}, 
\left.\frac{\delta}{m}\right|_{\times m},0\right),
\label{d2}\end{equation}
where 
$$
m\geq \delta^\frac{q}{q-1}\left(1+{q \delta} n^{1-q}\right)^\frac{1}{1-q}. 
$$

Then \(\| r_\delta-p \|_1=2\delta\), however
$$
\|\ p^q_\delta-p^q\ \|_1=((1-\delta)^q-1)n^{1-q} + \delta^q  m^{1-q}\geq
1.
$$
\hfill \bsq

Since the logarithm and the identity are injective continuous functions, we have:

{\bf Corollary 2.1} The Rényi and Tsallis entropies, if $q<1$, are not
continuous.\bsq

Note that the proof of the previous proposition is essentially identical 
that of Lesche in \cite{Les04a}, regarding
the instability of Rényi entropy. However,  the above argumentation 
is not applicable in the case $1<q$. In particular, it is impossible to determine $m$ so
that \(m^{1-q}\geq\delta^{-q}(1+(1-(1-\delta)^q)n^{1-q}   \)), because then the 
direction of the inequality is reversed by the negative powers 
$$
m^{q-1}\leq \frac{\delta^{q}}{(1+(1-(1-\delta)^q)n^{1-q})}<\delta^{q}.
$$

Hence, there is no $m\in \mathbb{N}$ that satisfies the inequality, if \(\delta<1\). 

Let us know investigate the continuity of the expectation values. Here it
is not enough to show that the function $p\mapsto \frac{p^q}{\|p^q\|}$ is
not continuous, because the strong convergence (convergence in norm) does
not follow from the weak convergence. What we show is that $p\mapsto (A|
p^q/\|p^q\|_1)$ is not continuous for a large set of $A\in l^\infty$.

Let be $p$ and $p^q$ are chosen as previously. Then 
\begin{equation}
 \|p^q_\delta\|_1 =(1-\delta)^{q}n^{1-q}+\delta^q m^{1-q}. 
\label{rnorm}\end{equation}
and  
$$
\frac{p^q_\delta}{\|p^q_\delta\|_1 }-\frac{p^q}{\|p^q\|_1 } =
\frac{\delta^{q}}{m^{q-1}\|p^q_\delta\|_1 }\left(\left.\left. -\frac{1}{n}\right|_{\times
n}, \frac{1}{m}\right|_{\times m}, 0\right).
$$

Therefore
\begin{gather}
\left|\left(A\left|,\frac{p^q_\delta}{\|p^q_\delta\|_1 }-
        \frac{p^q}{\|p^q\|_1 }\right. \right)\right|=
\frac{\delta^{q}}{m^{q-1}\|p^q_\delta\|_1 } 
        \left| -\frac{1}{n} \sum_{i=1}^n A_{i}+
        \frac{1}{m}\sum_{i=n+1}^{n+m} A_{i}\right| \nonumber\\
=\frac{\delta^{q}}{(1-\delta)^{q}\left(\frac{n}{m}\right)^{1-q}+\delta^q} 
        \left| -\frac{1}{n} \sum_{i=1}^n A_{i}+
        \frac{1}{m}\sum_{i=n+1}^{n+m} A_{i}\right|. 
\label{lim}\end{gather}

This expression is convergent as $m$ goes to infinity with the following
limit:\ $$
L:= \left| -\frac{1}{n} \sum_{i=1}^n A_{i}+
       \bar A_{(n)}\right|,
$$
where $\bar A_{(n)}=\lim_{m\rightarrow \infty} \frac{1}{m}\sum_{i=n+1}^{n+m} A_{i}$.
If $L$ is not zero - and it is not zero for most $A$-s - then we can choose an $m$
\ so  that \re{lim} is greater than $L/2$. Therefore we have proved, that

{\bf Proposition 3.} 
If $q<1$, then the q-expectation value of $A\in l^\infty$, $D_q\rightarrow \mathbb R, 
\ p\mapsto (A \ |\ p^q/\|p^q\|)\) is not continuous if $A$ satisfies
is an $n\in \mathbb{N}$ so that   
$$
\left| -\frac{1}{n} \sum_{i=1}^n A_{i}+
       \lim_{m\rightarrow \infty} \frac{1}{m}\sum_{i=n+1}^{n+m} A_{i}\right|
       \neq 0.
$$
\bsq

Note that a number of $A$-s satisfy this condition.

\section{Lesche stability and continuity}

The original mathematical formulation of experimental robustness by Lesche
is  not continuity, but a related notion. He introduced "normalized" values of
the corresponding functions instead of the "bare" values in the above definition
of continuity \cite{Bas05a,Abe05a}. To clarify the relation of Lesche
stability and continuity we introduce some additional notions. Let
us see then the following sets 
$$
V_n := \{p\in D\ |\ p_i=0\ \text{if}\ \ i>n\} \qquad n\in \mathbb{N},
$$
$$
V:=\bigcup_n V_n.
$$

It is clear that $V_m \in V_n$ if $m\leq n$. If $p\in V$ then let us define
$$
n_p := \min\{n\in \mathbb{N}\ |\ p\in V_n\}.
$$

Hence $p_i=0$ if $i>n_p$. 

Let $f:V\rightarrow \mathbb{R}$, \(f\neq 0\) be a function with the property 
\begin{displaymath}
\kappa_n:=\sup\{|f(p)|\ |\  p\in V_n\} < \infty \qquad (n\in \mathbb{N}).
\end{displaymath}

It is evident, that \(\kappa_m \leq \kappa_n\), if $m\leq n$. Moreover, there
is an $n_0\in \mathbb{N}$ so that $\kappa_{n_0}>0$.

\textit{Definition 2}: A function $f$ with the previous properties is Lesche-stable, if
$$
(\forall \epsilon>0)(\exists \delta>0)(\forall n>n_0)
        (\forall r,p\in V_n)\left(\|r-p\|<\delta\Rightarrow 
        \frac{|f(r)-f(p)|}{\kappa_n}< \epsilon\right),
$$ 
or equivalently
$$
(\forall \epsilon>0)(\exists \delta>0)
        (\forall r,p\in V)\left(\|r-p\|<\delta\Rightarrow 
        \frac{|f(r)-f(p)|}{\kappa_n}< \epsilon\right),
$$
where $n:=\max \{n_q,n_p\}>n_0$.

This definition corresponds to Lesche's original formulation \cite{Les82a}.

Comparing the definitions of continuity and Lesche-stability it is clear,
that
\begin{enumerate}
\item If $f$ is uniformly continuous, then it is Lesche-stable.
\item  If $f$ is bounded and Lesche-stable, then it is uniformly continuous.
\end{enumerate}

Lesche stability is a \emph{global} property, However, the physical meaning of
experimental robustness requires a refinement which is a \emph{local} property.  For
example let us see the intuitive formulation of experimental robustness of Abe \cite{Abe08a}:
\begin{quote}
"Given a statistical mechanical system, perform a measurement to obtain a
probability distribution $\{p_i\}_{i=1,...,w}$ ... . Perform a measurement
again on the same system prepared in the same state as before. Then another
probability distribution  $\{p'_i\}_{i=1,...,w}$ will be obtained."
\end{quote}

Continuing Abe requires that some related physical quantities do not be very
different. 

This formulation indicates that we want that in
the neighbourhood of an \textit{arbitrarily given} state the related physics
do not change dramatically. The uniformity does not seem to be important.

Therefore we introduce the following concept of stability.

\emph{Definition 3:} A function  $f$ is \emph{stable at $p\in V$} if
$$
(\forall \epsilon>0)(\exists \delta>0)(\forall r\in V\ \text{and}\
n_r>n_0)\left(\|r-p\|<\delta\Rightarrow 
        \frac{|f(r)-f(p)|}{\kappa_{n_{r}}}< \epsilon\right).
$$ 
A function $f$ is \emph{stable} if it is stable at all states of its domain. Lesche-stability is uniform stability.

It is easy to see, that
\begin{enumerate}
\item If $f$ is continuous in $p$, then it is stable there.
\item If $f$ is bounded and stable in $p$, then it is continuous there.
\item If $f$ is Lesche-stable, then it is stable everywhere.
\item If $f$ is stable in a compact set of its domain, then it is Lesche-stable there.
\item If $f$ is instable then it is also Lesche-instable.
\end{enumerate}

\section{The stability of Rényi and Tsallis entropies and q-expectation values}

\subsection{1<q} We have shown in section \ref{stab1}, that the Rényi and
Tsallis entropies are everywhere continuous, therefore they are stable. 

We have also seen that the q-expectation value of a physical quantity $A\in
l^\infty$ is continuous everywhere, therefore the q-expectation
value is stable.

If $A\notin l^\infty$, then the q-expectation value is not necessarily continuous,
however, it is stable. 

In this case
$$
  \kappa_n= \sup\left\{\left| \sum_{i=1}^n A_i\frac{p_i^q}{\|p^q\|}\right|\ \left|\
  p\in V_n\right.\right\} = \max_{i\leq n} |A_i|,
$$
therefore, if $n:= \max\{n_r, n_p\}>n_0$, then
$$
  \frac{\left| \sum_{i=1}^n A_i \left(\frac{r_i^q}{\|r^q\|} - \frac{p_i^q}{\|p^q\|}\right)\right|}{\kappa_{n_r}}
  \leq \frac{\kappa_{n_p}}{\kappa_{n_0}} \sum_{i=1}^n  \left|\frac{r_i^q}{\|r^q\|} - \frac{p_i^q}{\|p^q\|}\right|.
$$
The second term at the right hand side of this inequality is the difference
of the continuous function $p\rightarrow  p^q/\|p^q\|_1$ at values $p$
and $r$, as we have seen in \ref{stab1}.
Therefore, the right hand side of this inequality is smaller than $\epsilon$
choosing an \(r\) closer to \(p\) than \(\delta={\kappa_{n_0}}/{\kappa_{n_p}} \epsilon\).

\subsection{q<1} 

In subsection \ref{nstab1} we have seen the Rényi and Tsallis entropies and
the q-expectation values are not continuous, now we will show that they are
not stable. 

We can check that by a simple modification of the proofs in \ref{nstab1}.

Considering $p$ in \re{d1} and $r=r_\delta$ in \re{d2} we get for the Rényi
entropy, that 
$$
\kappa_{n_r}^{\text{Rényi}} = \log(n+m)
$$ 
and therefore the stability criteria
is\begin{equation}
\frac{|S_R(r)-S_R(p)|}{\kappa^{\text{Rényi}}_{n_{r}}} =
\frac{\log((1-\delta)^qn^{1-q}+\delta^qm^{1-q})-\log n^{1-q}}{\log(n+m)}.
\label{scR}\end{equation}

This expression converges to $1-q$ as $m$ goes to infinity.
Therefore the Rényi entropy is instable.    

Similarly for the Tsallis entropy we get 
$$
\kappa_{n_r}^{\text{Tsallis}} =\frac{ n^{1-q}+m^{1-q}-1}{1-q},
$$
and the stability criteria is
\begin{equation}
\frac{|S_T(r)-S_T(p)|}{\kappa^{\text{Tsallis}}_{n_{r}}} =(1-q)
\frac{|(1-(1-\delta)^q)n^{1-q}-\delta^qm^{1-q}|}{ n^{1-q}+m^{1-q}-1}.
\label{scT}\end{equation}

This expression is convergent as $m$ goes to infinity and has the limit
$$
0<L_{T} = \frac{1-q}{1-n^{q-1}}(1-(1-\delta)^q).
$$
Therefore choosing $m$ so that \re{scT} be greater than $L_T/2$, we see
that the Tsallis entropy is instable.

Regarding the stability of q-expectation values, it is enough to investigate
only the case $A\in l^\infty$. Now 
$$
\kappa_{n_r}^{\text{q-av.}} \leq\|A\|_\infty,
$$ therefore the expression \re{lim} divided by $\|A\|_\infty$ estimates
the corresponding expression of the stability criteria. If $A$ has the property
given in Proposition 3 then the q-averages are instable.  

\section{Discussion}

We have investigated some possible mathematical formulations of the experimental
robustness of some physical quantities. The analysis of continuity, uniform continuity,
and Lesche-stability revealed that these notions are closely related and
it is convenient to introduce to use a local stability concept instead of
the uniform notion of Lesche-stability. These formulations give essentially the same conditions of experimental robustness
for the   investigated functions:

The Rényi and Tsallis entropies  are continuous and stable if $1<q$  and are
not continuous and instable for finite uniform distributions, if $q<1$. 

The
q-expectation values are continuous and stable if \(A\in l^\infty\) and \(1<q\)\ 
 and are not necessarily  continuous but stable if \(A\notin l^\infty\) and \(1<q\). 
  The
q-expectation values are not continuous and instable for practically all  physical
 quantities \(A\in l^\infty\) (see the condition in \ref{nstab1}) in case of finite uniform
distributions.  

Observe that the proof of Lesche \cite{Les82a} and Abe \cite{Abe08a} for Lesche instability in the case
in the case \(1<q\) does not negate our stability because they do not consider a neighbourhood of a   
\textit{given} distribution (e.g. formula (7) in \cite{Abe08a}) but a sequence of  finite
distributions whose length goes to infinity.
The proof of Abe works in the case \(q<1\) but it shows the instability only for a single distribution.

If \(f\) is stable on a compact set of its domain, then it is also Lesche-stable.
If \(f\) is instable then it is also Lesche-instable.

\bibliographystyle{unsrt}

\end{document}